\documentclass[
    aps,prb,twocolumn,
	groupedaddress,superscriptaddress,
	amsfonts,amssymb,amsmath,
	citeautoscript,longbibliography,
	letterpaper, nofootinbib
	]{revtex4-2}

\usepackage[utf8]{inputenc}
\usepackage[english]{babel}

\usepackage{microtype} %For better kerning and symbol-stretching
\usepackage{xspace} %For the \xspace command

\usepackage{txfonts}  %Times Roman fonts
\usepackage{txfontsb} %Addition for txfonts, including old style numerals and greek

\usepackage{bm} %Bold math symbols with \bm{} (Greek and other symbols)

\usepackage{xcolor}
\usepackage[]{graphicx} % "demo" option to disable rendering for speed
\graphicspath{{figs/}}

\usepackage[]{booktabs}
\usepackage{array}
\usepackage{layouts}
\usepackage{multirow}

\usepackage{enumerate}
\usepackage[inline]{enumitem}

\usepackage{xr}
\makeatletter
\newcommand*{\addFileDependency}[1]{% argument=file name and extension
  \typeout{(#1)}% latexmk will find this if $recorder=0 (however, in that case, it will ignore #1 if it is a .aux or .pdf file etc and it exists! if it doesn't exist, it will appear in the list of dependents regardless)
  \@addtofilelist{#1}% if you want it to appear in \listfiles, not really necessary and latexmk doesn't use this
  \IfFileExists{#1}{}{\typeout{No file #1.}}% latexmk will find this message if #1 doesn't exist (yet)
}
\makeatother

\usepackage{hyperref}
\hypersetup{colorlinks,
	linkcolor={blue!75!black!80!yellow},
	citecolor={blue!75!black!80!yellow},
	urlcolor={blue!75!black!80!yellow}
}

\usepackage[capitalize,nameinlink]{cleveref}

\crefname{subequations}{Eqs.}{Eqs.} %Specific changes to allow for Eqs.-wording when referring to a set of subequations. Label of subequations must include [subequations] as an option.
\Crefname{subequations}{Eqs.}{Eqs.}
\crefformat{subequations}{#2Eqs.~(#1)#3}
\Crefformat{subequations}{#2Eqs.~(#1)#3}
\crefname{page}{p.}{p.} %Changing from 'page' to 'p.'
\crefname{table}{Table}{Tables}
\crefname{figure}{Figure}{Figures}
\crefname{section}{Section}{Sections}

\usepackage{placeins}

\usepackage{siunitx}
\DeclareSIUnit[number-unit-product = ]\percent{\char`\%} % remove spacing for \percent

\usepackage[centering,hmargin=18mm,tmargin=29.4mm,bmargin=24mm]{geometry}

\usepackage{soul}

\usepackage{textcomp} % for \textrightarrow
\usepackage{xifthen}
\usepackage{etoolbox}
\newboolean{togglecomments}
\newboolean{toggletodos}
\newboolean{togglechanges}

\setboolean{togglecomments}{true}
\setboolean{toggletodos}{true}
\setboolean{togglechanges}{false} 

\newcommand{\textblacksquare}{$\blacksquare$}
\newcommand{\todo}[1]{\ifbool{toggletodos}%
	{\textcolor{green!60!black}{\small\textsf{{}\textsuperscript{\textsc{\textsf{todo}}}}[\ignorespaces#1]}} % if true, show comments
	{}}     % if false, do nothing
\newcommand{\comment}[2]{\ifbool{togglecomments}%
		{\textcolor{blue!70!black}{\small\sf\textsuperscript{\textsc{\textsf{\ignorespaces#1}}}[\ignorespaces#2]}} % if true, show comments
		{}}     % if false, do nothing
\newcommand{\swap}[2]{\ifbool{togglechanges}
	{\ignorespaces#2}  % revisions-only version
	{\textcolor{red!70!black}{[\ignorespaces#1]}\textrightarrow{}\textcolor{green!50!black}{[\ignorespaces#2]}}}
\newcommand{\remove}[1]{\ifbool{togglechanges}
	{}    % revisions-only version
	{\textcolor{blue}{\ignorespaces#1}}}
\newcommand{\inset}[1]{\ifbool{togglechanges}
	{\ignorespaces#1}  % revisions-only version
	{\textcolor{green!50!black}{\ignorespaces#1}}}

\newcommand{\citeremind}[1]{%
	[\textcolor{blue!75!black!80!yellow}{\textblacksquare%
		\ifthenelse{\isempty{#1}}{}{\textsuperscript{\tiny\textsf{\ignorespaces#1}}}%
	}]\xspace}

\newcommand{\ie}{i.e.,\@\xspace} %Gobble-spaces of the "small" type (small is ensured by adding \@)

\newcommand{\appropto}{\mathrel{\vcenter{
			\offinterlineskip\halign{\hfil$##$\cr
				\propto\cr\noalign{\kern.2pt}\sim\cr\noalign{\kern-2.5pt}}}}}

\DeclareFontFamily{U}{mathx}{\hyphenchar\font45}
\DeclareFontShape{U}{mathx}{m}{n}{<5> <6> <7> <8> <9> <10>
                                  <10.95> <12> <14.4> <17.28> <20.74> <24.88>
                                  mathx10}{}
\DeclareSymbolFont{mathx}{U}{mathx}{m}{n}
\DeclareFontSubstitution{U}{mathx}{m}{n}
\makeatletter
\newcommand{\raisemath}[1]{\mathpalette{\raisem@th{#1}}}
\newcommand{\raisem@th}[3]{\raisebox{#1}{$#2#3$}}
\makeatother

\renewcommand{\paragraph}[1]{\vskip 1ex\noindent\textbf{#1.}~}

\usepackage{braket}
\usepackage[eulergreek]{sansmath}
\makeatletter
\renewcommand\@make@capt@title[2]{%
    \@ifx@empty\float@link{\@firstofone}{\expandafter\href\expandafter{\float@link}}%
    \sisetup{math-sf=\textsf}%
    \sansmath\sffamily\textbf{#1\@caption@fignum@sep}#2 % does not work with the newtx* packages unfortunately
}%

\makeatother

\graphicspath{{figures/}}
\setboolean{togglecomments}{true}
\setboolean{toggletodos}{true}
\setboolean{togglechanges}{true} 
\begin{document}
\title{Refining Heuristic Predictors of Fractional Chern Insulators using Machine Learning}

\author{Oriol Mayné i Comas$^\bigstar$}
\affiliation{Department of Physics, Massachusetts Institute of Technology, Cambridge, Massachusetts 02139, USA}
\affiliation{Universitat Polit\`ecnica de Catalunya, Barcelona 08034, Spain}

\author{André Grossi Fonseca$^\bigstar$}
\email{agfons@mit.edu}
\affiliation{Department of Physics, Massachusetts Institute of Technology, Cambridge, Massachusetts 02139, USA}
\affiliation{The NSF Institute for Artificial Intelligence and Fundamental Interactions}

\author{Sachin Vaidya}
\email{svaidya1@mit.edu}
\affiliation{Department of Physics, Massachusetts Institute of Technology, Cambridge, Massachusetts 02139, USA}
\affiliation{The NSF Institute for Artificial Intelligence and Fundamental Interactions}
\affiliation{Research Laboratory of Electronics, Massachusetts Institute of Technology, Cambridge, Massachusetts 02139, USA\\
$^\bigstar$ denotes equal contribution}

\author{Marin Solja\v ci\'c}
\affiliation{Department of Physics, Massachusetts Institute of Technology, Cambridge, Massachusetts 02139, USA}
\affiliation{The NSF Institute for Artificial Intelligence and Fundamental Interactions}
\affiliation{Research Laboratory of Electronics, Massachusetts Institute of Technology, Cambridge, Massachusetts 02139, USA\\
$^\bigstar$ denotes equal contribution}

\begin{abstract}
We develop an interpretable, data-driven framework to quantify how single-particle band geometry governs the stability of fractional Chern insulators (FCIs). 
Using large-scale exact diagonalization, we evaluate an FCI metric that yields a continuous spectral measure of FCI stability across parameter space. 
We then train Kolmogorov–Arnold networks (KANs)—a recently developed interpretable neural architecture—to regress this metric from two band-geometric descriptors: the trace violation $T$ and the Berry curvature fluctuations $\sigma_B$. 
Applied to spinless fermions at filling $\nu=1/3$ in models on the checkerboard and kagome lattices, our approach yields compact analytical formulas that predict FCI stability with over $>80 \%$ accuracy in both regression and classification tasks, and remain reliable even in data-scarce regimes.
The learned relations reveal model-dependent trends, clarifying the limits of Landau-level-mimicking heuristics. 
Our framework provides a general method for extracting simple, phenomenological “laws” that connect many-body phase stability to chosen physical descriptors, enabling rapid hypothesis formation and targeted design of quantum phases.
\end{abstract}
\maketitle 

\section{Introduction}

Fractional quantum Hall (FQH) physics constitutes a striking example of emergence and topological order in many-body physics. 
In two-dimensional (2D) electron gases with partially filled lowest Landau level (LL), the holomorphic structure of single-particle wavefunctions enables precise analytical descriptions, such as ansatz wavefunctions~\cite{laughlin_anomalous_1983} and exact pseudopotential arguments~\cite{haldane1983pseudopotential, pokrovskySimpleModelFractional1985, trugman1985exact}.
The discovery of fractional Chern insulators (FCIs)~\cite{parameswaran2013fractional,bergholtz_topological_2013,liu_recent_2024,neupert_fci_2011,sun_2011_flatband, tang_high-temperature_2011,Sheng2011,regnault_fractional_2011, hafezi_fractional_2007, wang_fractional_2011} extended this physics to lattice systems without magnetic fields, where topologically ordered states arise from partially filled Chern bands. Recent experiments in moiré superlattices~\cite{spanton2018observation, xie2021fractional, park_observation_2023, zeng2023thermodynamic,lu_fractional_2024, xie2025tunable, lu2025extended, aronson2025displacement, cai2023signatures, xu2023fci, ji2024local, redekop2024direct, xu2025signatures, wu2025modeling} have provided compelling evidence for the existence of such states.

However, partially filling a nearly flat Chern band alone does not guarantee an FCI ground state. 
Recently, the role of single-particle band geometry in stabilizing FCIs has been elucidated~\cite{parameswaran2013fractional,liu_recent_2024,roy_band_2014,jackson_geometric_2015,Lee2017,Ledwith2020,ozawa_relations_2021,Mera2021b,Wang2021,varjas2022,ledwith_vortexability_2023,Estienne2023,Okuma, shi2025effects}. 
In particular, two geometric quantities have been found to be the most relevant: the ``trace violation'' $T$, which quantifies deviations from the lowest LL relation between the Fubini–Study metric and Berry curvature; and the Brillouin zone fluctuations of the Berry curvature $\sigma_B$. 
These metrics quantify similarities to the algebra of projected density operators in the lowest LL~\cite{roy_band_2014, girvin_magneto-roton_1986}, and their single-particle nature means they are computationally efficient surrogates for costly many‐body computations.
In practice, however, these quantities vary together, making it difficult to disentangle  their individual roles or assess their predictive power. 
%While several case studies have offered insight into their influence on FCI stability, a general framework for disentangling and comparing these effects has remained elusive. 
This lack of quantitative understanding limits both the interpretation of numerical results and the ability to design systems optimized for fractionalized ground states.

In this work, we take a step toward filling this gap by developing a data-driven approach to evaluate how well these geometric heuristics predict FCI stability.
%We use high-throughput exact diagonalization to evaluate a recently devised ``target-phase loss function'', which quantitatively assesses the stability of an FCI.
We use high-throughput exact diagonalization (ED) to evaluate the FCI quality~\cite{fonseca2025gradient} across parameter space, which is a continuous metric quantifying the stability of, or distance from, an FCI phase.
We then employ Kolmogorov–Arnold networks (KANs)~\cite{liu2024kan}—a symbolic, interpretable machine-learning (ML) architecture—to extract simple and accurate symbolic formulas, akin to phenomenological ``laws'' for how this quality metric depends on band geometry.
We apply our pipeline to two prototypical tight‐binding models for spinless fermions on the checkerboard and kagome lattices at filling $\nu = 1/3$.
Strikingly, we find that simple formulas involving only $T$ and $\sigma_B$ predict FCI stability with over $80\%$ accuracy in both regression and classification tasks. 
The learned relations also reveal model-dependent behavior: while large $\sigma_B$ destabilizes the kagome FCI, it enhances stability in the checkerboard lattice, highlighting the limitations of heuristics that solely mimic LLs. Moreover, our approach remains suitable even in data-scarce regimes, accurately capturing these trends with as few as $\sim 10^2$ ED samples.
Our results offer an interpretable framework for evaluating the stability of many-body quantum phases as a function of arbitrary physical quantities of interest. 

This paper is organized as follows.
In Section II, we describe the dataset, define the FCI quality metric, and detail the lattice models and band-geometric predictors.
Section III reviews Kolmogorov–Arnold networks (KANs) and outlines the evaluation pipeline used to extract symbolic formulas.
Section IV presents our main results, including representative formulas, model-dependent trends, and their physical interpretation.
Finally, Section V concludes with discussions of our results and outlook.

\section{Background and dataset}

\subsection{Exact diagonalization and FCIs}

In the exact diagonalization (ED) method, one directly constructs and computes the low-energy spectrum of a many-body Hamiltonian $H(\mathbf{p})$, typically a function of physical parameters $\mathbf{p}$. 
Numerical efficiency can be achieved by exploiting symmetries to block-diagonalize $H(\mathbf{p})$, in which each block corresponds to a symmetry sector and can therefore be labeled by quantum numbers $\mathcal{K}_i$.
For example, in a 2D system with discrete translational symmetry, the components of the center-of-mass momentum along each primitive lattice vector are good quantum numbers.
Consequently, the Hamiltonian can be constructed and diagonalized within each irreducible representation of the translation group. 
A typical ED spectrum then consists of eigenenergies as a function of discrete quantum numbers, $\{ \boldsymbol{E}_{\mathcal{K}_i} (\mathbf{p}) \}$.

Although the Hilbert‐space dimension within each sector grows exponentially with system size, this symmetry resolution greatly reduces the computational cost and provides insights into the ground state and excitation spectra, making ED an indispensable tool for strongly correlated systems. 
Due to the physics of FCIs being typically dominated by short-range correlations~\cite{laughlin_anomalous_1983, haldane1983pseudopotential, pokrovskySimpleModelFractional1985, trugman1985exact}, FCI ground states can be resolved at modest system sizes, and therefore ED has been extensively employed to study their behavior. 
In finite clusters at fractional fillings, FCIs are identified in ED through a combination of signatures: a quasi-degenerate ground-state manifold separated by a clear spectral gap, a characteristic spectral flow under adiabatic flux insertion, and an entanglement gap and a peculiar level count in the particle-cut entanglement spectrum~\cite{bergholtz_topological_2013, liu_recent_2024}. 
Based on this numerical evidence, ED has been used to identify FCI ground states in a variety of lattices~\cite{neupert_fci_2011,Sheng2011,regnault_fractional_2011, wu_zoology_2012, hafezi_fractional_2007, wang_fractional_2011}, as well as in models in the continuum~\cite{abouelkomsan2020, repellin2020, Lanchli21, crepel2023, liu2021, li2021, reddy2023fci}.

\subsection{FCI quality metric}

We now introduce a metric that turns the low-energy spectrum $\{ \boldsymbol{E}_{\mathcal{K}_i} (\mathbf{p}) \}$ of an ED calculation into a continuously varying stability measure for FCIs.
For this, we employ a modified version of the recently introduced ``target-phase loss function''~\cite{fonseca2025gradient}, which we now briefly review.

Being topologically ordered phases, FCIs are known to display ground state degeneracy on a torus, which becomes a quasi-degeneracy due to finite-size splitting. 
The ground-state manifold can then be labeled by a set of symmetry sectors $\{\mathcal{K}_i^*\}$, with corresponding degeneracies $\{d_i^*\}$.
For a particular FCI, the data $\{(\mathcal{K}_i^*, d_i^*)\}$ is known beforehand for any system size with $N_1 \times N_2$ unit cells along primitive lattice vectors~\cite{regnault_fractional_2011, bernevig_emergent_2012, ardonne2008degeneracy}. 
Then, we split $\{ \boldsymbol{E}_{\mathcal{K}_i} (\mathbf{p}) \}$ into two sets, which we call the target and complement manifolds.
The former consists of the lowest $d_i^*$ energy levels in each $\mathcal{K}_i^*$ sector, and the latter comprises all remaining energy levels. 
We now introduce the function
\begin{equation}
\label{eq:loss_per_flux}
    \tilde{\ell}(\mathbf{p}, \Phi) = \max_{E \in\, \operatorname{target}}E(\mathbf{p}, \Phi) - \min_{E \in\, \operatorname{complement}}E(\mathbf{p}, \Phi),
\end{equation}

which, at a given threaded magnetic flux $\Phi$, is simply the gap between the ground state energy of the complement manifold and the highest excited state energy in the target manifold. 
If $\tilde{\ell}(\mathbf{p}, \Phi) < 0$, then the system at point $\mathbf{p}$ and flux $\Phi$ is in the target phase at $\mathbf{p}$ with associated many-body gap $-\tilde{\ell}$. 
If $\tilde{\ell}(\mathbf{p}, \Phi) > 0$, the loss function encodes the spectral distance from an FCI ground state.
We then ensure that the many-body gap stays open under adiabatic flux insertion~\cite{oshikawa_commens_2000} by defining
\begin{equation}
\label{eq:general_loss}
    \tilde{\mathcal{L}}(\mathbf{p}) = \max_{\{ \Phi_j \}} \tilde{\ell}(\mathbf{p}, \Phi_j),
\end{equation}
where the maximum is taken over a finite user-defined sampling of flux values $\{ \Phi_j \}$. 
Since the $\max$ function is negative if and only if all of its arguments are negative, $\tilde{\mathcal{L}}(\mathbf{p}) < 0$ signals that the many-body gap remains finite for every flux in the set $\{ \Phi_j \}$.

These definitions provide an initial basis for defining the quality of ($\tilde{\mathcal{L}} < 0$) or distance from ($\tilde{\mathcal{L}} > 0$) an FCI ground state.
However, the regime in which $\tilde{\ell}(\mathbf{p}, \Phi) < 0$ only contains information about the many-body gap. 
In practice, the quality of an FCI should also account for the energy spread of the ground state quasi-degeneracy, defined as
\begin{equation}
\label{eq:spread}
    \delta(\mathbf{p}, \Phi) = \max_{E \in\, \operatorname{target}}E(\mathbf{p}, \Phi) - \min_{E \in\, \operatorname{target}}E(\mathbf{p}, \Phi),
\end{equation}
with smaller values of $\delta(\mathbf{p}, \Phi)$ denoting more stable FCIs.
However, note that it is not enough to simply take the ratio of \cref{eq:loss_per_flux} and \cref{eq:spread}, since $\delta(\mathbf{p}, \Phi)$ is not meaningful in the case where $\tilde{\ell}(\mathbf{p}, \Phi) > 0$.
To ensure $\delta(\mathbf{p}, \Phi)$ is accounted for only when the ground state is an FCI, we divide $\tilde{\ell}$ by a function which smoothly interpolates between $\delta$ and unity when $\tilde{\ell}(\mathbf{p}, \Phi)$ goes from negative to positive.
In particular, we choose
\begin{equation}
\label{eq:modified_loss_per_flux}
    \ell(\mathbf{p}, \Phi) = \frac{\tilde{\ell}(\mathbf{p}, \Phi)}{ \delta(\mathbf{p}, \Phi) + (1-\delta(\mathbf{p}, \Phi))\, \sigma(\tilde{\ell}(\mathbf{p}, \Phi)/\delta(\mathbf{p}, \Phi))},
\end{equation}

where $\sigma(x) = (1+\exp(-x))^{-1}$ is the sigmoid function.
Since $\sigma(x) \rightarrow 1, 0$ for $x \rightarrow \pm \infty$,
$\ell(\mathbf{p}, \Phi) \rightarrow \tilde{\ell}(\mathbf{p}, \Phi)$ for large positive $\tilde{\ell}(\mathbf{p}, \Phi)$, but $\ell(\mathbf{p}, \Phi) \rightarrow \tilde{\ell}(\mathbf{p}, \Phi)/\delta(\mathbf{p}, \Phi)$ for large negative $\tilde{\ell}(\mathbf{p}, \Phi)$, on the scale of the spread, as desired. 
Finally, we define the FCI quality metric to be
\begin{equation}
\label{eq:modified_general_loss}
    \mathcal{L}(\mathbf{p}) = \max_{\{ \Phi_j \}} \ell(\mathbf{p}, \Phi_j).
\end{equation}
Note that the division by $\delta(\mathbf{p}, \Phi)$ in the case $\tilde{\ell}(\mathbf{p}, \Phi) < 0$ means that datasets are typically imbalanced, with negative values of $\mathcal{L}(\mathbf{p})$ showing absolute values much larger than those points with positive $\mathcal{L}(\mathbf{p})$, which is suboptimal for ML methods.
We address this imbalance in Section III.

\subsection{Models}

Here, we focus on two well-known tight-binding models that have been shown to host FCIs, the first on the checkerboard lattice~\cite{regnault_fractional_2011} and the second on the kagome lattice~\cite{wu_zoology_2012}.
Their varying numbers of sublattices and connectivities produce distinct band geometries and interaction matrix elements, making them ideal for exploring universal versus model-specific FCI behavior.

\begin{figure}
    \centering
    \includegraphics[width=\linewidth]{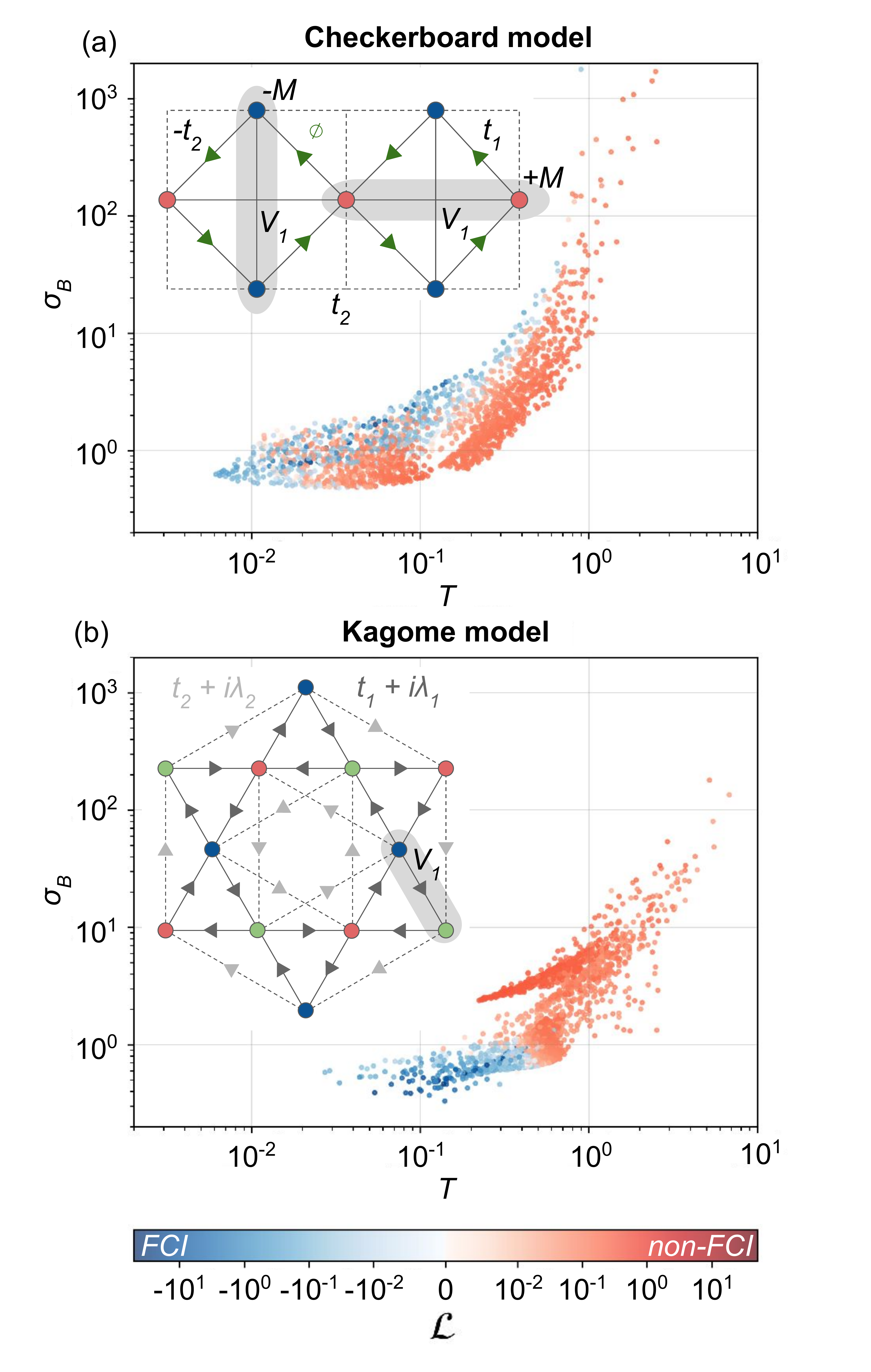}
    \caption{Log-log plots of the dataset used for ML analysis, representing the distribution of trace violation $T$ and Berry curvature deviation $\sigma_B$, colored by the corresponding FCI quality metric $\mathcal{L}$. 
    Data is generated from ED calculations using the Hamiltonian on (a) the checkerboard lattice (\cref{eq:checkerboard_model} and (b) kagome lattice (\cref{eq:kagome_model}). 
    Insets show a schematic representation of the lattice models, with arrows indicating hoppings with positive phase.}
    \label{fig:scatterplot}
\end{figure}

The Hamiltonian we consider on the checkerboard lattice is
\begin{equation}
\label{eq:checkerboard_model}
\begin{split}
&H = H_{\mathrm{kin}} + H_{\mathrm{int}}, \\
&H_{\mathrm{kin}} = -\sum_{\langle i,j\rangle} t_{1} \exp(i\nu_{ij}\phi) c_{i}^{\dagger}c_{j} 
- \sum_{\langle\langle i,j\rangle\rangle}t_{2}\nu_{ij} c_{i}^{\dagger}c_{j} + \sum_i M \mu_i \,n_i, \\
&H_{\mathrm{int}} = V_{1}\sum_{\langle i,j\rangle}:\!n_{i}\,n_{j}\!:,
\end{split}
\end{equation}
where $c_i\, (c_i^\dag)$ is the fermion annihilation (creation) operator at site $i$, $n_i = c_i^\dag c_i$ is the density operator, and $:\,\,:$ indicates normal ordering.
The nearest-neighbor (NN) and next-nearest-neighbor (NNN) sites are represented by $\langle i,j\rangle, \langle\langle i,j\rangle\rangle$.
$t_1$ and $\phi$ parameterize the absolute value and phase of the NN hopping; $t_2$ parameterizes the purely real NNN hopping, with $\nu_{ij}=\pm 1$ encoding a chosen chirality for complex hoppings, as shown in \cref{fig:scatterplot}a; $M$ is a mass term with alternating sign $\mu_i = \pm 1$ for different sublattices.
Interactions are included through repulsive NN density-density terms with strength $V_1$.
Throughout, we set $V_1=1$ to fix the energy scale.  

The second model, on the kagome lattice, has the same interaction Hamiltonian $H_\mathrm{int}$ as above but with the kinetic Hamiltonian given by
\begin{equation}
\label{eq:kagome_model}
\begin{split}
&H_{\mathrm{kin}} = -\sum_{\langle i,j\rangle} (t_{1}+i\nu_{ij}\lambda_1) c_{i}^{\dagger}c_{j} 
- \sum_{\langle\langle i,j\rangle\rangle}(t_{2}+i\nu_{ij}\lambda_2)c_{i}^{\dagger}c_{j},
\end{split}
\end{equation}
where $t_1$ and $\lambda_1$ parameterize the real and imaginary parts of the NN hopping, $t_2$ and $\lambda_2$ parameterize the real and imaginary parts of the NNN hopping, and $\nu_{ij}=\pm 1$ is a chirality chosen in the same convention as \cref{fig:scatterplot}b. 
As in the checkerboard model, we fix $V_1  = 1$.

Since we are entirely focused on the role of band geometry, we remove energy dispersion effects by flattening the lowest single-particle band and projecting the interactions to it, which effectively amounts to sending single‐particle gaps to infinity without altering the band geometry~\cite{bergholtz_topological_2013}. 
In this projected regime, only the single‐particle eigenstates enter the many‐body calculation, so one may also set $t_{1} = 1$ for both models. 
Consequently, the parameter spaces are three-dimensional (3D), $\mathbf{p}_{CB}=(\phi,t_{2},M)$ for the checkerboard model and $\mathbf{p}_{KM}=(\lambda_{1},t_{2},\lambda_{2})$ for the kagome model.
Previous work~\cite{regnault_fractional_2011, wu_zoology_2012} has shown that the points around $\mathbf{p}_{CB} =(\pi/4,1-\sqrt{2}^{-1},0)$ and $\mathbf{p}_{KM} =(1,0,0)$ host FCI ground states.
All ED calculations in our dataset are carried out at filling $\nu=1/3$ of the lowest band. 
We use clusters with sizes $N_1 = 3, N_2 = 6$, since the resulting momentum grid includes all high-symmetry momentum points, and therefore our calculations can resolve competing orders such as charge density waves. 
For evaluations of the FCI quality metric, we sample the two extremal flux values $\{ \Phi_j \} = \{ 0, \pi \}$.

\subsection{Single-particle metrics}

As inputs to our ML pipeline, we calculate two single-particle band-geometric quantities on the partially filled lowest band: the trace violation $T$ between the Fubini--Study metric and Berry curvature, and the Berry curvature fluctuations $\sigma_B$ across the Brillouin zone (BZ).
The quantum geometric tensor is

\begin{equation}
\label{eq:qgt}
    \eta_n^{\mu\nu}(\mathbf{k}) = \sum_n \bra{\partial_{k_\mu}u_n(\mathbf{k})}Q(\mathbf{k})\ket{\partial_{k_\nu}u_n(\mathbf{k})}, 
\end{equation}

where $u_n(\mathbf{k})$ is a cell-periodic Bloch state in band $n$ and $Q(\mathbf{k}) = 1 - \sum_n \ket{u_n(\mathbf{k})}\bra{u_n(\mathbf{k})}$.
Its real and imaginary parts are the Fubini--Study metric and Berry curvature, respectively:

\begin{equation}
\label{eq:fsmetric+bc}
    g_n^{\mu\nu}(\mathbf{k}) = \operatorname{Re} \eta_n^{\mu\nu}(\mathbf{k}), \quad \mathcal{F}_n(\mathbf{k}) \varepsilon^{\mu\nu} = \frac{1}{2} \operatorname{Im} \eta_n^{\mu\nu}(\mathbf{k}),
\end{equation}

with $\varepsilon^{\mu\nu}$ the 2D Levi--Civita symbol.
Finally, by focusing on the lowest band, the single-particle metrics of interest are defined as follows:

\begin{equation}
\label{eq:spmetrics}
\begin{split}
    &T = \frac{1}{2\pi} \int_{\operatorname{BZ}} d^2k\, T(\mathbf{k}), \quad T(\mathbf{k}) = \operatorname{Tr}g^{\mu\nu}(\mathbf{k}) - |\mathcal{F}(\mathbf{k})|, \\   
    &\sigma_B = \sqrt{\int_{\operatorname{BZ}} \frac{d^2k}{A_{BZ}}\left(\frac{\mathcal{F}(\mathbf{k})}{\mathcal{F}_0} - C\right)^2\,}.
\end{split}
\end{equation}

where $A_{BZ}$ is the BZ area, $\mathcal{F}_{BZ} = 2\pi/A_{BZ}$ and $C$ the Chern number of the lowest band.
Both quantities are normalized to be dimensionless and vanish for the lowest LL.
We also include the band gap $\Delta$ as an input, which controls band mixing when compared to interaction scales and the bandwidth.
However, because we work in the flattened projected limit, $\Delta$ is effectively sent to infinity as discussed previously, and therefore is irrelevant for the FCI stability.
We shall see below that, although $\Delta$ is treated as an input, our ML method is able to detect its irrelevance.

Motivated by the algebra of projected density operators in LLs~\cite{girvin_magneto-roton_1986, roy_band_2014}, these geometric metrics are expected to serve as efficient surrogates for FCI stability, although in some cases their role has been challenged~\cite{simon_fractional_2015, varjas2022, ledwith_vortexability_2023, yang_singular_2025, lin_fractional_2025, fonseca2025gradient}. 
The assembled dataset of paired geometric inputs and FCI quality labels $\{(T, \sigma_B, \Delta; \mathcal{L})\}$ constitute the dataset for our ML analysis.
We note that, in general, the FCI quality metric cannot be expressed as a sole function of any collections of single-particle quantities.  
Our goal here instead is to quantitatively and systematically assess to what extent are $T$ and $\sigma_B$ reliable predictors of an FCI ground state in the parameter regions we study, and derive heuristic laws describing their relationship.

For the dataset, we focus on the behavior of $\mathcal{L}$ near and within a region of parameter space where the ground state is an FCI.
Therefore, we generate the data for each model by sampling parameter points in a 3-sphere centered at $\mathbf{p}_{CB}$ and $\mathbf{p}_{KM}$, with radii chosen such that there is a roughly equal number of FCIs and non-FCIs in the dataset, \ie points with $\mathcal{L}(\mathbf{p}) < 0$ and $\mathcal{L}(\mathbf{p}) > 0$, respectively.
In \cref{fig:scatterplot} we plot the datasets used for each model as a function of $T$ and $\sigma_B$.

\section{Kolmogorov--Arnold networks}

\subsection{Introduction}

Neural networks (NNs) are a class of ML models designed to approximate complex, nonlinear mappings between inputs and outputs. 
The predictive capability of a NN arises from multiple compositions of linear transformations and nonlinear activation functions, the depth of which relates to its expressivity, \ie the complexity of functions it can approximate. 
During training, the coefficients of these transformations are adjusted in order to minimize a user-defined loss function, which quantifies the discrepancy between predicted outputs and ground-truth values.
Training typically proceeds via gradient-based optimization: the gradient of the loss with respect to each parameter is computed using backpropagation, and the parameters are iteratively updated in a direction that reduces the loss. 
This procedure enables the network to adapt its internal representation to capture the underlying functional relationship in the data.

Among the most fundamental architectures are multilayer perceptrons (MLPs). 
An MLP consists of a sequence of layers, each comprising multiple nodes. 
Denoting the inputs and outputs of the $j$-th node of a layer as $\mathbf{x}$ and $y_j$, the computation performed at a node can be represented as:
\begin{equation}
    y_j(\mathbf{x}) = \psi\left( \sum_i w_{ij} x_i + b_j \right),
\end{equation}
where $w_{ij}, \,b_j$ are the node's weights and bias and $\psi(x)$ is a \emph{fixed} nonlinear activation function. 
By stacking multiple such layers, an MLP is able to construct increasingly abstract feature representations, making it capable of approximating arbitrary continuous functions.

\begin{figure}
    \centering
    \includegraphics[width=\linewidth]{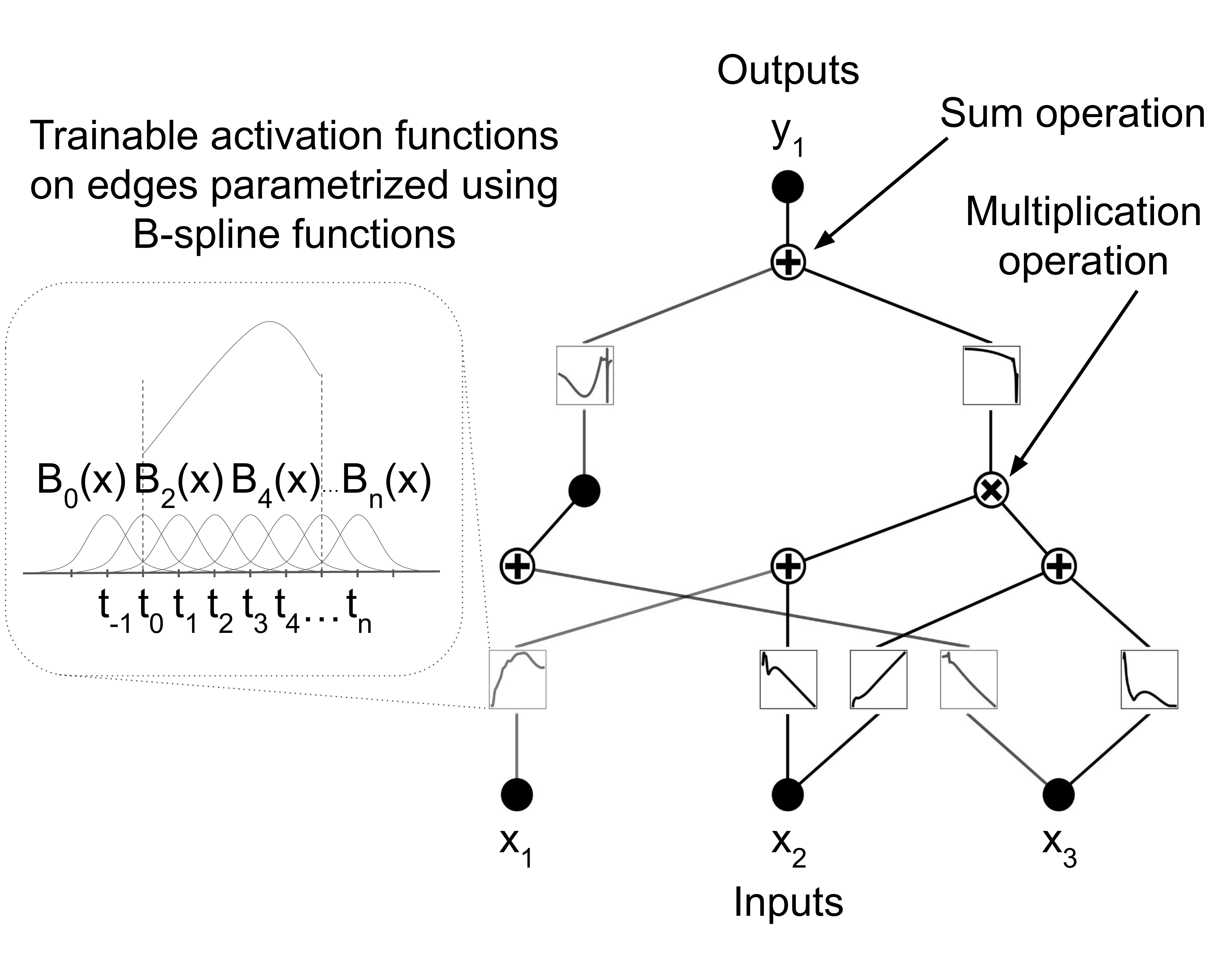}
    \caption{Example of a KAN [3,[1,1],1] architecture. 
    In KAN notation, an architecture with $n$ layers is described by a list of $n$ entries, with the $i$-th entry specifying the number of addition nodes in the $i$-th layer.
     Layers with two numbers (i.e. [1,1]) indicate the number of additive and multiplicative nodes at that layer, respectively.
    }
    \label{fig:demo_kan}
\end{figure}

On the other hand, Kolmogorov--Arnold networks (KANs)~\cite{liu2024kan} are a novel neural network architecture based on the Kolmogorov--Arnold representation theorem, which states that, under mild assumptions, a continuous multi-variable function $f(\mathbf{x})$ can be represented as a finite composition of continuous single-variable functions and additions:
\begin{equation}
    f(\textbf{x}) = \sum_{q} \Phi_{q} \left( \sum_p \phi_{q,p}(x_p) \right),
\end{equation}
which can be seen as a 2-layer network parametrized by single-variable functions ($\Phi_q$, $\phi_{q,p}$). 
However, in general such functions can be pathological, which disallows gradient-based methods for function approximation. 
The KAN architecture instead employs multiple layers of smooth single-variable functions, which are then added or multiplied together at each node, and composed between layers as shown in~\cref{fig:demo_kan}.
%Each node performs a linear combination of transformations of the previous layer’s outputs using the nonlinear transformations $\phi_i$ in its incoming edges
%\begin{equation}
%    f(\boldsymbol{x}) = \sum_{i=1}^{n_{in}} \phi_i(x_i)
%\end{equation}

Intuitively, KANs can be understood as generalizations of MLPs, where the nonlinear activation functions are \emph{trainable} rather than fixed.
The activation functions $\psi(x)$ are parametrized as a combination of B-splines $B_i(x)$:

\begin{equation}
    \psi(x) =\omega_b\frac{x}{1 + e^{-x}} +  \omega_s\sum_ic_iB_i(x),
    \label{eq:edge_formula}
\end{equation}

where the parameters $c_i$, $\omega_s$ and $\omega_b$ can be trained with standard backpropagation methods to fit any nonlinear relation. 
%In essence, the $c_i$ parameters adjust the weight of each B-spline $B_i(x)$ to conform the overall spline function to its desired shape, while $\omega_s$ and $\omega_b$ adjust the magnitude of the basis function and the spline function in relation to one another. 

The use of KANs in this work is motivated by their noticeable advantages over MLPs in terms of compactness and interpretability. 
While the typically high connectivity of MLPs hinder the extraction of general insights from the architecture, the trainable parameters in KANs are packaged within each edge rather than through increasing the network complexity, often leading to simpler networks for a given accuracy. 
Furthermore, KANs are trained with sparsity enforcement in the loss function followed by pruning, in which unnecessary edges and nodes are iteratively dropped from the network to reduce its size and complexity. 
This allows for fitting simple, single-variable functions to the nonlinear activation functions after training. 
This fitting procedure is typically done by choosing functions for each edge from a predetermined library of candidates, which can be compared through their $R^2$ values.
Therefore, KANs allow for the integration of both NN-based predictions and symbolic regression into a single model, making it particularly useful for this problem, achieving both accurate prediction of $\mathcal{L}$ as well as simple, interpretable symbolic formulas for its approximate dependence on $(T, \sigma_B, \Delta)$.
The full training process, which includes many design choices such as sparsity tuning, optimizer selection, specific training loss, hyperparameter tuning and pruning schedule, is crucial for achieving optimal results and is discussed in detail below and in Appendix A.

In order to evaluate the model performance on our FCI dataset, we use the root mean square error (RMSE) and accuracy, defined as
 \begin{equation}
 \label{eq:ml_metrics}
\begin{split}
     &\text{RMSE} =\sqrt{\frac{1}{n}\sum_{i=1}^n(y_i - \hat{y_i})^2}, \\
     &\text{Accuracy} = \frac{\text{Correctly classified samples}}{\text{Total number of samples}},
\end{split}
\end{equation}
where $\hat{y}_i$ and $y_i$ denote the ground-truth and predicted labels.
%These statistical metrics measure the regression and classification performance of the model, respectively.
``Correctly classified sample" is defined as the number of samples in which the sign of the FCI metric $\mathcal{L}$ is predicted correctly by the KAN model.
%Other metrics, such as recall and precision, are also used when examining a model in detail, and are discussed further in the Appendix \comment{AGF}{We never actually discuss this in the Appendix, so I am in favor of just dropping this last part}. \comment{SV}{In that case we can drop it. I do remember Oriol showing these metrics for the final models. If we have them, we can consider including them.}

The clear data trends in \cref{fig:scatterplot}, 
%with separate clusters for both FCI and non-FCI classes related to the values of the variables. 
%Both datasets reveal a strongly non-linear relationship between the input data and the FCI quality metric, making them particularly suitable for modeling with KANs, given that they can model the dependencies directly (using the trainable edges) rather than relying on compositionality, as would be the case in MLPs. 
with distinct clusters for FCI and non-FCIs visible and reasonably low overlap, suggest that KAN models should be able to distinguish between the two classes using the input variables and simple symbolic functions alone.

%Another important advantage of KANs is that networks tend to be noticeably more compact than their MLP counterparts for a given number of parameters. 
%This is because many of the trainable parameters in KANs are packaged within each edge rather than through increasing the network complexity. 
%Therefore, trained KANs lead to compact, accurate formulas, from which we can easily infer the overall heuristics learned by the models.

\subsection{Pipeline for formula extraction}

We use a comprehensive pipeline to integrate the model training, evaluation, formula selection and final evaluation. 
The steps are the following, which are depicted visually in \cref{fig:pipeline_diagram}a and expanded upon in Appendix A:

\begin{enumerate}
    \item \textbf{Hyperparameter selection:} based on iterative tests and experimentation, a comprehensive set of hyperparameter values, including a range of possible architecture sizes, are selected to train multiple candidate KANs. 
    These hyperparameters are chosen according to their performance in terms of prediction loss and accuracy, while favoring those architectures that offer the greatest network simplicity in terms of connectivity and compactness.
    
    \item \textbf{Model evaluation:} once the models are trained, the best networks are chosen, with performance characterized by a balance between low RMSE, high accuracy and low network size.
    
    \item \textbf{Candidate formula generation:} based on the selected models and a predetermined library of candidate functions, each of the models is adjusted to multiple formulas, with each edge having a high $R^2$ fit coefficient. 
    The list of candidate formulas is then chosen to ensure high performance with the fewest and simplest candidate functions.
    
    \item \textbf{Formula evaluation:} once the formulas are selected, we evaluate them on the regression and classification metrics.
    %accounting for the complexity of the formulas selected (measured quantitatively in terms of the $R^2$ product of all edges as well as qualitatively). 
    Out of the many possible candidates, a final representative for each model is selected, based on performance and formula simplicity.
\end{enumerate}

We shall see that KANs are able to meet these stringent requirements for high accuracy, low network size and high sparsity, allowing for physically meaningful insights to be extracted from the final architectures through symbolic regression.
In \cref{fig:pipeline_diagram}b we show illustrative KAN networks obtained after training, which capture the dependence of $\mathcal{L}$ on band geometry. 

\begin{figure}
    \centering
    \includegraphics[width=\linewidth]{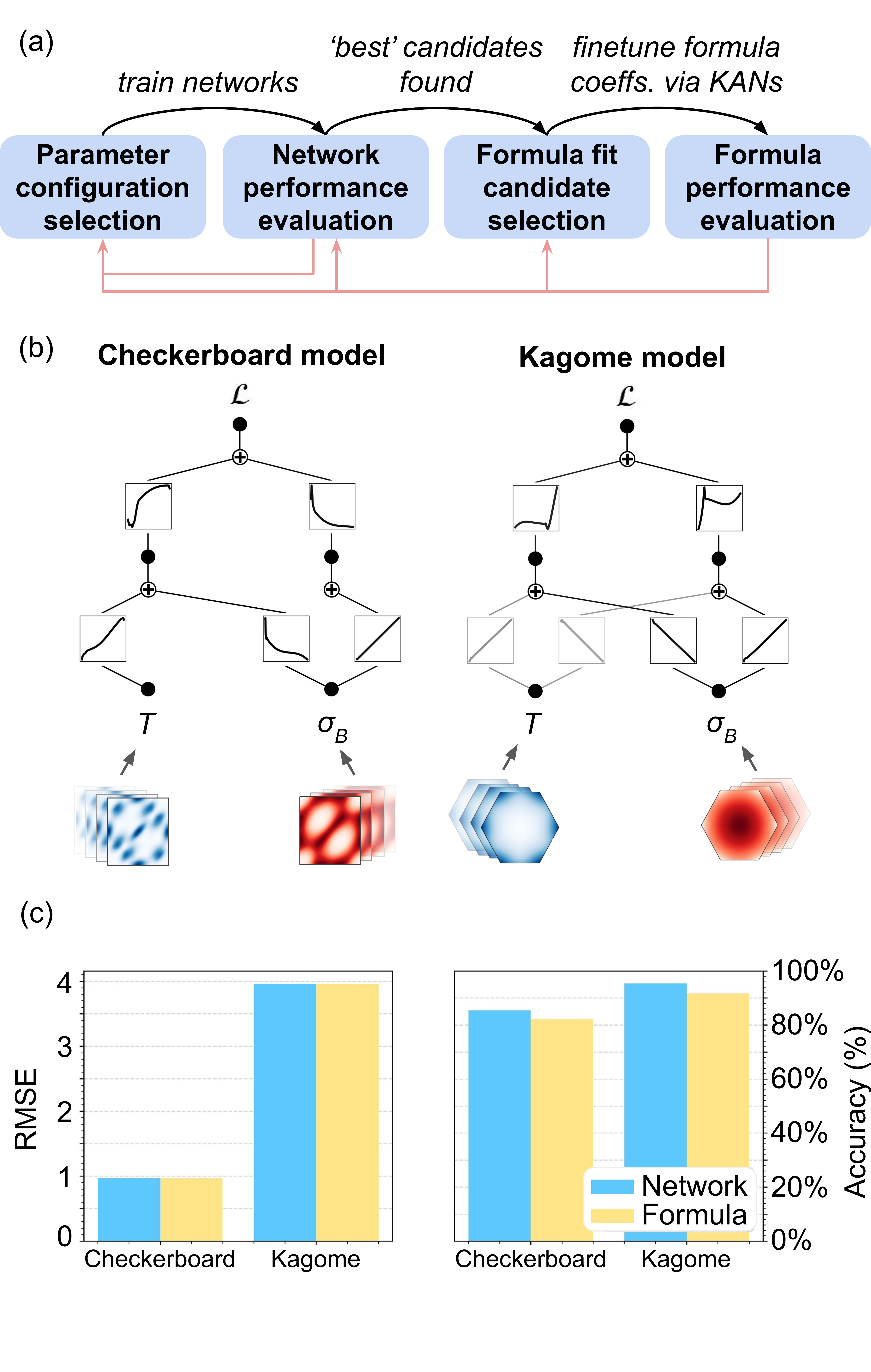}
    \caption{(a) ML pipeline for training KANs and extracting accurate symbolic formulas.
    (b) Representative KAN networks for the checkerboard and kagome models.
    The momentum-space functions $T(\mathbf{k})$ and $\mathcal{F}(\mathbf{k})$ (plotted respectively in blue and red, in each Brillouin zone) are used to compute $T$ and $\sigma_B$, which are then fed into the networks as inputs, with output the FCI metric $\mathcal{L}$.
    (c) Regression and classification performance of networks and symbolic formulas, measured in terms of the RMSE and accuracy, respectively. 
    }
    \label{fig:pipeline_diagram}
\end{figure}

\section{Results}

% \comment{OM}{The idea behind this idea I put was that the relation is simple enough to be 'seen' more or less with the naked eye, and thus a few non-linear functions would be enough to actually get a great formula, which meant that an interpretable KAN could be extracted; this is both in opposition to MLPs, which might require a relatively large and non-interpretable network to achieve through compositionality what the KAN can do in a layer or two, and also in opposition to very complex non-linear relationships that need many layers, in which case the KAN is no longer as interpretable and might not be as worth it}

On training KANs on our FCI dataset, we first find that KANs effectively distinguish between relevant and irrelevant variables, systematically dropping the single-particle gap $\Delta$ whenever it was included as an input parameter.
As discussed previously, in a band-projected ED calculation, any band mixing effects are completely removed, and therefore $\Delta$ plays no direct role in determining the ground state, being relevant only insofar as it affects the distribution of $T(\mathbf{k})$ and $\mathcal{F}(\mathbf{k})$, which is already captured by $T$ and $\sigma_B$.

\cref{table:tab1} presents representative formulas and associated coefficients for each of the models.
Although all of the expressions obtained are simple and interpretable, they are also accurate both in regression and classification metrics, as shown in \cref{fig:pipeline_diagram}c.
In particular, the KAN models and symbolic formulas for both lattice models scored above 80\% in accuracy, even reaching above 90\% for the kagome model.
This is a testament to both the suitability of our KAN pipeline, as well as the predictive power of band-geometric properties in the likelihood and stability of an FCI ground state.
The fact that the RMSEs show greater discrepancy between the models can be explained by noting the wider range of values of $\mathcal{L}$ in the kagome model (\cref{fig:scatterplot}), which leads to larger errors due to outliers that amplify the overall mean. 
This aspect was partially mitigated with a sigmoid transformation performed within the training pipeline, which is detailed in Appendix A.

\begin{table*}[t]
\centering
\begin{tabular}{|c|c|c|}
\hline
\textbf{Model} & \textbf{Formula} & \textbf{Params} \\ \hline
  Checkerboard &  $\mathcal{L}(T, \sigma_B) = - w_0 + w_1 T + (w_2 \sigma_B + w_3)^{-1}$ & $w_0 = 0.461$, \, $w_1 = 0.765$, \, $w_2 = 0.454$, \, $w_3 = 2.010$  \\ \hline
   Kagome &  $\mathcal{L}(T, \sigma_B) = w_0 - w_1 \exp{(-w_2 T)} - w_3 \exp{(-w_4\sigma_B)}$  & $w_0 = 0.112$, \, $w_1 = 0.175$, \, $w_2 = 4.109$, \, $w_3 = 0.311$, \,  $w_4 = 1.542$   \\ \hline
\end{tabular}
\caption{Representative formulas for the FCI metric $\mathcal{L}$ as a function of single-particle band geometric quantities, with learned parameters $w_i$ also shown.
}
\label{table:tab1}
\end{table*}

%The following list presents the final formulas extracted from the models:

%\begin{itemize}
%    \item \textbf{Checkerboard formula:} $- 0.461 + 0.765T + \frac{1.840}{0.835\sigma_B + 3.699}$
%    \item \textbf{Kagome formula:} $0.112 - 0.311e^{-1.542\sigma_B} - 0.175e^{-4.109T}$
%\end{itemize}

%For the checkerboard lattice, the overlap is quite noticeable, with a significant region of the parameter space hosting samples from both classes. In the Kagome lattice, conversely, separation is significantly clearer, with a sharp boundary and minimal overlap that suggest a potentially higher accuracy down the line. 

Overall, these formulas allow us to take the usual FCI heuristics one step further and quantitatively evaluate general trends of the FCI quality for each model.
In the formula for the checkerboard model, $\mathcal{L}$ is lowered, thereby resulting in a more stable FCI, if $T$ is decreased but if $\sigma_B$ is \emph{increased}, although the functional dependencies are such that $\mathcal{L}$ is far more sensitive to changes in $T$ than to in $\sigma_B$. 
In contrast, the formula for the kagome model shows the same exponential dependence on both variables, such that an FCI ground state is favored if both $T$ and $\sigma_B$ are decreased. 
Although the specific functional dependencies on band geometry are formula-dependent and therefore not universal, in Appendix B we show that the general trends discussed above are consistent across formulas.
The behavior of the checkerboard model, favoring increases in $\sigma_B$ for more stable FCIs, is in contrast to the traditional expectation for FCIs, but it aligns with more recent work which has pointed out that a low $\sigma_B$ can be irrelevant~\cite{ledwith_vortexability_2023} or even detrimental~\cite{varjas2022} to the stability of an FCI.
Here, we provide evidence for and partially recover these results in a purely data-driven manner.

\begin{figure}
    \centering
    \includegraphics[width=\linewidth]{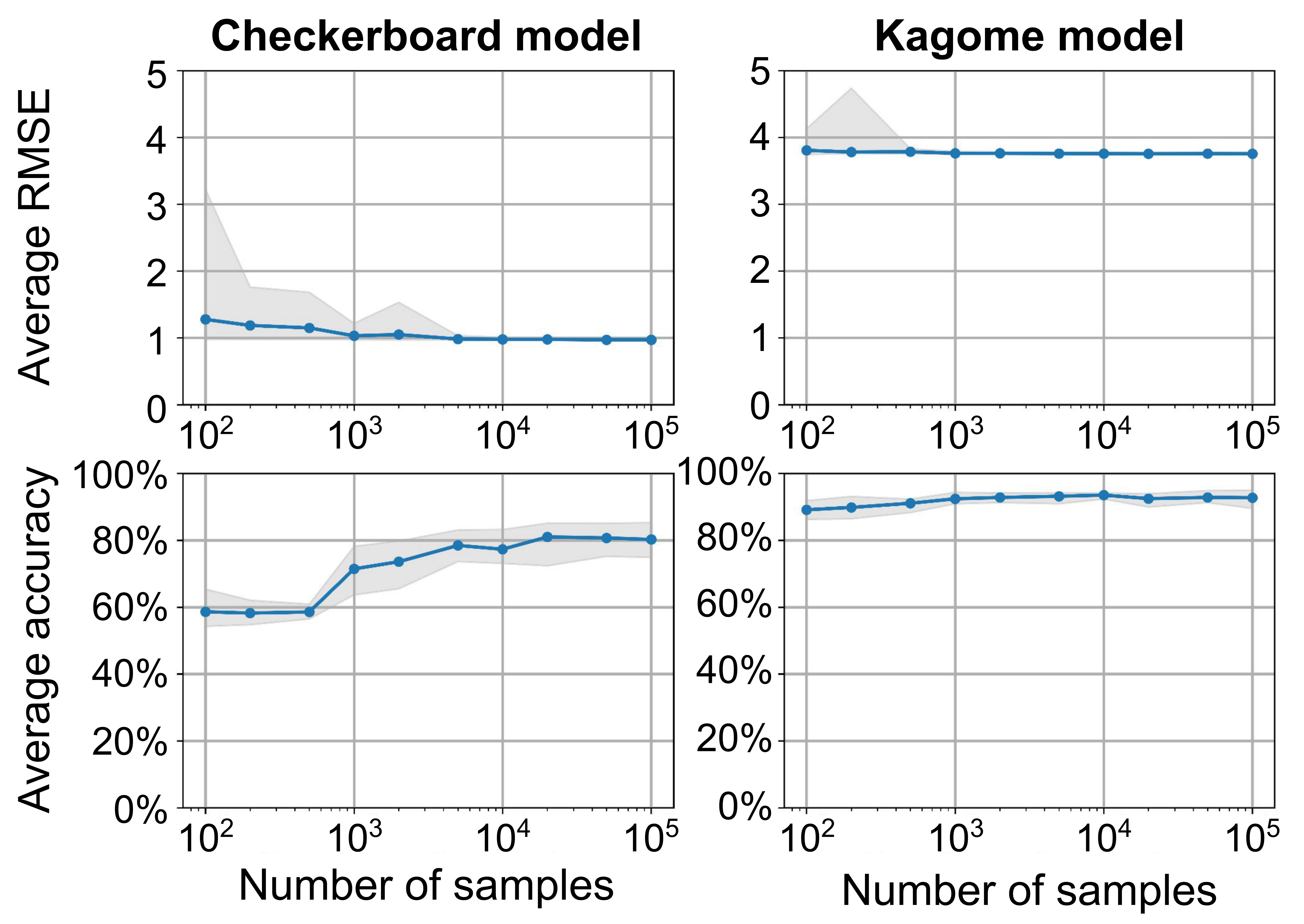}
    \caption{Average RMSEs and accuracies as a function of training dataset size, with error bars in grey.
    }
    \label{fig:number_samples_performance}
\end{figure}

Finally, the intrinsic exponential complexity of ED calculations typically hinders the generation of sizable, high-quality datasets.
This motivates us assess the performance of KANs in data-scarce scenarios. 
We do so by training KANs on a small set of sampled points with the same distribution of FCIs and non-FCIs, and testing the performance of the final network and symbolic expressions on unseen data. 
As shown in \cref{fig:number_samples_performance}, the regression and classification metrics remain robust for small datasets, with performance on the kagome model largely unchanged down to as few as $\sim 10^{2}$ ED spectra. 
These trends demonstrate that KANs retain strong predictive power in data-scarce settings, delivering stable statistical metrics without requiring a large number of data points.

\section{Conclusions}

In this work, we have addressed a key challenge in the study of fractional Chern insulators (FCIs) using a data-driven approach: developing interpretable and quantitative heuristics that link single-particle band geometry to FCI stability. 
To achieve this, we combined large-scale exact diagonalization with a physically motivated FCI metric $\mathcal{L}$~\cite{fonseca2025gradient} and interpretable machine learning via Kolmogorov–Arnold networks (KANs)~\cite{liu2024kan}. 
This pipeline allowed us to extract symbolic formulas that connect the trace violation $T$ and Berry curvature fluctuations $\sigma_B$ to FCI stability.

Our results show that remarkably simple formulas for $\mathcal{L}$ involving only $T$ and $\sigma_B$ can predict whether an FCI is the ground state of the system and quantitatively capture its stability, with high accuracy across two representative models on the checkerboard and kagome lattices. 
Both models exhibit strong nonlinear but learnable relationships in the $(T, \sigma_B)$ plane, with the extracted formulas achieving over 80\% classification accuracy and competitive regression performance. 
Interestingly, we find model-dependent trends: while large $\sigma_B$ is detrimental to stability in the kagome model, it plays a stabilizing role in the checkerboard lattice, in line with recent theoretical results~\cite{ledwith_vortexability_2023, varjas2022}. 
Therefore, KANs were able to surface unusual trends in the stability dependence on $\sigma_B$ in a purely data-driven approach, without the need for fine-tuned models where $T$ and $\sigma_B$ can be separately controlled.
The extracted symbolic expressions also offer insights into the ground state of these quantum systems and the crucial role of single-particle physics, delivering accurate and predictive results even for small-scale datasets, a crucial requirement in numerical many-body physics.

Looking ahead, this framework provides a general methodology for exploring and discovering complex dependencies of many-body states on arbitrary physical quantities.
Our findings illustrate the unique potential of KANs and ML in accelerating exploration in the physical sciences through hypothesis formulation and testing.

\section{Acknowledgements}
We thank Zhuo Chen and Patrick J. Ledwith for stimulating discussions.
We acknowledge support from the National Science Foundation under Cooperative Agreement PHY-2019786 (The NSF AI Institute for Artificial Intelligence and Fundamental Interactions). 
O.M.C.\ acknowledges the CFIS Mobility Program for its support, particularly Fundació Privada Mir-Puig, CFIS partners, and donors of the crowdfunding program.
A.G.F.\ acknowledges support from the Whiteman Fellowship and the Surpina and Panos Eurnekian Nanotechnology Fund Fellowship. 
S.V. and M.S.\ acknowledge support from the U.S.\ Office of Naval Research (ONR) Multidisciplinary University Research Initiative (MURI) under Grant No.\ N00014-20-1-2325 on Robust Photonic Materials with Higher-Order Topological Protection.
This material is based upon work also supported in part by the U. S. Army Research Office through the Institute for Soldier Nanotechnologies at MIT, under Collaborative Agreement Number W911NF-23-2-0121.
The MIT SuperCloud and Lincoln Laboratory Supercomputing Center provided computing resources that contributed to the results reported in this work.

\section{Appendix A: training pipeline details}
\label{app:training_pipeline}

\begin{table*}[!ht]
\centering
\begin{tabular}{|c|c|c|c|c|}
\hline
\textbf{Model} & \textbf{Formula} & \textbf{Params} & \textbf{RMSE} & \textbf{Acc.} \\ \hline
  Checkerboard &  $\mathcal{L} = w_0 - (w_1 T + w_2)^{-1} + (w_3 \sigma_B + w_4)^{-1}$ & $w_0 = 1.157$, \, $w_1 = 0.317$, \, $w_2 = 0.617$, \, $w_3 = 0.489$, \, $w_4 = 2.007$ & 0.98 & 83.19 \% \\ & $\mathcal{L} = w_0 - (w_1 T + w_2)^{-1} + w_3 \exp{(-w_4 \sigma_B)}$ & $w_0 = 1.237$, \, $w_1 = 0.321$, \, $w_2 = 0.617$, \, $w_3 = 0.399$, \, $w_4 = 0.194$ & 0.98 & 81.78 \% \\ \hline
   Kagome &  $\mathcal{L} = w_0 - w_1 \exp{(-w_2 T) - (w_3 \sigma_B - w_4)^{-1}}$ & $w_0 = 0.129$, \, $w_1 = 0.175$, \, $w_2 = 4.091$, \, $w_3 = 12.967$, \, $w_4 = 0.913$ & 3.91 & 91.66 \%   \\ \hline
\end{tabular}
\caption{Additional formulas for the FCI metric $\mathcal{L}$ as a function of single-particle band geometric quantities, and associated learned parameters $w_i$, RMSE and accuracies.
}
\label{table:tab2}
\end{table*}

Here we detail the training process, including choices of optimizer, hyperparameters and initial KAN architectures.
As is typical in ML problems, such choices are essential to achieve optimal model performance. 
The following configuration was used to train the final models within PyTorch:

\begin{itemize}
    \item \textbf{AdamW optimizer} with the default configuration.
    \item \textbf{smoothL1 loss} over $N$ samples: $\mathcal{L} = \frac{1}{N}\sum_{i=1}^NL(\hat{y_i},y_i)$, where 
    \[
    L(\hat{y},y) = 
    \begin{cases}
    0.5(\hat{y} - y)^2, & \text{if } |\hat{y} - y| < 1 \\
    |\hat{y} - y| - 0.5, & \text{otherwise}
    \end{cases}
    \]
    
    \item \textbf{120 training epochs} with pruning every 20 epochs.
    \item \textbf{Exponential decay scheduler} with starting  learning rate $lr = 5 \times 10^{-3}$ and weight decay $\gamma = 0.98$. 
\end{itemize}

For data sampling and transformation, we chose:

\begin{itemize}
    \item \textbf{$10^5$ training samples} extracted from each dataset, with $2 \times 10^4$ validation samples and $10^5$ test samples for a uniform final evaluation.
    \item \textbf{$10^3$ samples per batch}, for a total of $10^2$ batches per epoch. 
    \item \textbf{Metric transformation:} In order to address the wide data range of FCI metrics, we applied the transformation $\mathcal{L}' = 2 \, \sigma(a \mathcal{L}) - 1$, mapping the metric to the range $\mathcal{L}' \in (-1, 1)$. 
    This transformation was performed on the predicted and ground-truth values \textit{after} the prediction. After experimentation, the value for \textit{a = 20} was chosen for both the kagome and checkerboard models, based on an ablation study performed on the variable. 
    \item \textbf{Uniform sampling:} samples were binned according to their transformed value for the FCI metric into $10^2$ different bins, with a roughly equal amount of samples selected from each bin to avoid oversampling some regions of the $(T, \sigma_B)$ plane.
\end{itemize}

The KAN-specific choices of architecture and hyperparameters were:

\begin{itemize}
    \item \textbf{Architecture:} we initialized KANs using the following sizes: [2,1]; [2,2,1]; [2,5,1]; [2,[2,2],1]; [2,2,2,1]; [2,[2,2],[2,2],1]; [2,5,5,1]; [2,[2,2],[2,2],[2,2],1]; [2,10,1]; [2,10,10,1].
    \item \textbf{Sparsity configuration:} 3 configurations were tested, introducing sparsity to the network at different rates (see definition of coefficients in \cite{liu2024kan}):
    \begin{enumerate}
        \item $\lambda = 10^{-3}$, $\lambda_e = 0.1$, $\lambda_{coeffdif} = 0.1$, $\lambda_{l1} = 0.1$, $\lambda_{coef} = 0.01$
        \item $\lambda = 5\cdot10^{-3}$, $\lambda_e = 0.25$, $\lambda_{coeffdif} = 0.25$, $\lambda_{l1} = 0.2$, $\lambda_{coef} = 0.01$
        \item $\lambda = 10^{-2}$, $\lambda_e = 0.5$, $\lambda_{coeffdif} = 0.5$, $\lambda_{l1} = 0.4$, $\lambda_{coef} = 0.01$
    \end{enumerate}

    \item \textbf{KAN model parameters:} we initialized KANs with the following set of parameters:
    \begin{itemize}
        \item \textbf{Grid size:} 20
        \item \textbf{\textit{k}:} 3
        \item \textbf{grid$_\epsilon$:} 0.01
    \end{itemize}
\end{itemize}

\remove{
Another important choice for training KANs is sparsity enforcement. 
As discussed in the main text, this is done by adding a term to the training loss that grows with the number of nodes and edges in the architecture:

\begin{equation}
    l_{spars} = \lambda \left( \mu_1 \sum_{l=0}^{L-1}|\Phi_l|_1 + \mu_2 \sum_{l=0}^{L-1} S(\Phi_l) \right)
\end{equation}

with $\mu_1$, $\mu_2$ and $\lambda$ being used to control the relative weight of each half and the overall weight of the penalization, respectively. Each of the individual terms refer to the following:

\begin{equation}
|\Phi|_1 \equiv \sum_{i=1}^{n_\text{in}} \sum_{j=1}^{n_\text{out}} |\phi_{i,j}|_1
\end{equation}

\begin{equation}
S(\Phi) \equiv - \sum_{i=1}^{n_\text{in}} \sum_{j=1}^{n_\text{out}}
\frac{|\phi_{i,j}|_1}{|\Phi|_1}
\log \left( \frac{|\phi_{i,j}|_1}{|\Phi|_1} \right)
\end{equation}

\begin{equation}
|\phi|_1 \equiv \frac{1}{N_p} \sum_{s=1}^{N_p} \big| \phi\big(x^{(s)}\big) \big|
\end{equation}

The term $|\Phi|_1$ computing the overall L1 norm of the entire layer according to the individual L1 norms of each component ($|\phi|_1$), while $S(\Phi)$ is the entropy associated to it.

Evaluation was performed on both the regression and classification metrics. As explained in the main body, both the precise value and the binary class were considered valuable predictions by the network, which prompted their evaluation. The specific formulas used were the following (recall and precision being valid for both classes):

\begin{itemize}
    \item \textbf{RMSE:} $\sqrt{\frac{1}{n}\sum_{i=1}^n(y_i - \hat{y_i})^2}$
    \item \textbf{Accuracy:} $\frac{TP \space + \space TN}{TP \space + \space TN \space + \space FP \space + \space FN}$
    \item \textbf{Recall:} $\frac{TP}{TP \space + \space FN}$
    \item \textbf{Precision:} $\frac{TP}{TP \space + \space FP}$
\end{itemize}

When selecting models, the focus was on RMSE and accuracy, being those the two most representative of each type. In practice, models selected were those with the lowest values in either metric, and those with smaller networks were chosen over those with larger ones. This overall criteria was also used for evaluating formulas in the final pipeline step, swapping the size of the network for the overall complexity of the formula.
}
For the symbolic regression part, a list of possible candidate functions was selected to be used when fitting each edge. 
After multiple rounds of evaluation considering the complexity and performance of the different candidates, the final list contained the following single-variable functions: $x, x^2, x^3, \frac{1}{x}, e^x$ and $0$, which we have found to be sufficient for providing meaningful and accurate expressions. 

For each edge, these formulas were fitted to the trained non-linear activation by tuning the affine parameters, and then compared through their $R^2$ value, where
\begin{equation}
    R^2 = \frac{\sum_i (\hat{y_i} - \bar{y})^2}{\sum_i(y_i - \bar{y})^2}.
\end{equation}
A threshold of 0.95 for the $R^2$ values was used, with all candidates above that value considered for each edge.
If none were above the threshold, only the single best function was considered. 
All possible candidate combinations were considered in the final step, in which the formula quality was evaluated.

\section{Appendix B: additional metric formulas}

In \cref{table:tab2} we show additional symbolic formulas generated for each model through the KAN pipeline, along with their statistical metrics.
We note that, although the specific functions can vary depending on training details, they all show similar trends and dependencies.

\bibliography{references}

@article{haldane1983pseudopotential,
  title = {Fractional Quantization of the {Hall} Effect: A Hierarchy of Incompressible Quantum Fluid States},
  author = {Haldane, F. D. M.},
  journal = {Phys. Rev. Lett.},
  volume = {51},
  issue = {7},
  pages = {605--608},
  numpages = {0},
  year = {1983},
  month = {Aug},
  publisher = {American Physical Society},
  doi = {10.1103/PhysRevLett.51.605},
  url = {https://link.aps.org/doi/10.1103/PhysRevLett.51.605}
}

@article{trugman1985exact,
  title = {Exact results for the fractional quantum {Hall} effect with general interactions},
  author = {Trugman, S. A. and Kivelson, S.},
  journal = {Phys. Rev. B},
  volume = {31},
  issue = {8},
  pages = {5280--5284},
  numpages = {0},
  year = {1985},
  month = {Apr},
  publisher = {American Physical Society},
  doi = {10.1103/PhysRevB.31.5280},
  url = {https://link.aps.org/doi/10.1103/PhysRevB.31.5280}
}

@article{Lanchli21,
	author = {Wilhelm, Patrick and Lang, Thomas C. and L\"auchli, Andreas M.},
	doi = {10.1103/PhysRevB.103.125406},
	issue = {12},
	journal = {Phys. Rev. B},
	month = {Mar},
	numpages = {16},
	pages = {125406},
	publisher = {American Physical Society},
	title = {Interplay of fractional Chern insulator and charge density wave phases in twisted bilayer graphene},
	url = {https://link.aps.org/doi/10.1103/PhysRevB.103.125406},
	volume = {103},
	year = {2021}}

@article{Sheng2011,
	author = {Sheng, D. N. and Gu, Zheng-Cheng and Sun, Kai and Sheng, L.},
	date = {2011/07/12},
	doi = {10.1038/ncomms1380},
	id = {Sheng2011},
	isbn = {2041-1723},
	journal = {Nat. Commu.},
	number = {1},
	pages = {389},
	title = {Fractional quantum {Hall} effect in the absence of {Landau} levels},
	url = {https://doi.org/10.1038/ncomms1380},
	volume = {2},
	year = {2011}}

@article{Okuma,
  title = {Constructing vortex functions and basis states of Chern insulators: Ideal condition, inequality from index theorem, and coherentlike states on the von Neumann lattice},
  author = {Okuma, Nobuyuki},
  journal = {Phys. Rev. B},
  volume = {110},
  issue = {24},
  pages = {245112},
  numpages = {13},
  year = {2024},
  month = {Dec},
  publisher = {American Physical Society},
  doi = {10.1103/PhysRevB.110.245112},
  url = {https://link.aps.org/doi/10.1103/PhysRevB.110.245112}
}

@article{Estienne2023,
  title = {Ideal Chern bands as Landau levels in curved space},
  author = {Estienne, Benoit and Regnault, Nicolas and Cr\'epel, Valentin},
  journal = {Phys. Rev. Res.},
  volume = {5},
  issue = {3},
  pages = {L032048},
  numpages = {6},
  year = {2023},
  month = {Sep},
  publisher = {American Physical Society},
  doi = {10.1103/PhysRevResearch.5.L032048},
  url = {https://link.aps.org/doi/10.1103/PhysRevResearch.5.L032048}
}

@article{regnault_fractional_2011,
	title = {Fractional {Chern} {Insulator}},
	volume = {1},
	copyright = {http://creativecommons.org/licenses/by/3.0/},
	issn = {2160-3308},
	url = {https://link.aps.org/doi/10.1103/PhysRevX.1.021014},
	doi = {10.1103/PhysRevX.1.021014},
	number = {2},
	urldate = {2024-12-04},
	journal = {Phys. Rev. X},
	author = {Regnault, N. and Bernevig, B. Andrei},
	month = dec,
	year = {2011},
	pages = {021014},
}

@article{xie2025tunable,
  title={Tunable fractional {Chern} insulators in rhombohedral graphene superlattices},
  author={Xie, Jian and Huo, Zihao and Lu, Xin and Feng, Zuo and Zhang, Zaizhe and Wang, Wenxuan and Yang, Qiu and Watanabe, Kenji and Taniguchi, Takashi and Liu, Kaihui and others},
  journal={Nature Materials},
  pages={1--7},
  year={2025},
  publisher={Nature Publishing Group UK London}
}

@article{lu2025extended,
  title={Extended quantum anomalous {Hall} states in graphene/hBN moir{\'e} superlattices},
  author={Lu, Zhengguang and Han, Tonghang and Yao, Yuxuan and Hadjri, Zach and Yang, Jixiang and Seo, Junseok and Shi, Lihan and Ye, Shenyong and Watanabe, Kenji and Taniguchi, Takashi and others},
  journal={Nature},
  volume={637},
  number={8048},
  pages={1090--1095},
  year={2025},
  publisher={Nature Publishing Group UK London}
}

@article{aronson2025displacement,
  title={Displacement field-controlled fractional {Chern} insulators and charge density waves in a graphene/{hBN} moir{\'e} superlattice},
  author={Aronson, Samuel H and Han, Tonghang and Lu, Zhengguang and Yao, Yuxuan and Butler, Jackson P and Watanabe, Kenji and Taniguchi, Takashi and Ju, Long and Ashoori, Raymond C},
  journal={Physical Review X},
  volume={15},
  number={3},
  pages={031026},
  year={2025},
  publisher={APS}
}

@article{cai2023signatures,
  title={Signatures of fractional quantum anomalous Hall states in twisted {MoTe2}},
  author={Cai, Jiaqi and Anderson, Eric and Wang, Chong and Zhang, Xiaowei and Liu, Xiaoyu and Holtzmann, William and Zhang, Yinong and Fan, Fengren and Taniguchi, Takashi and Watanabe, Kenji and others},
  journal={Nature},
  volume={622},
  number={7981},
  pages={63--68},
  year={2023},
  publisher={Nature Publishing Group UK London}
}

@article{xu2023fci,
  title = {Observation of Integer and Fractional Quantum Anomalous {Hall} Effects in Twisted Bilayer ${\mathrm{MoTe}}_{2}$},
  author = {Xu, Fan and Sun, Zheng and Jia, Tongtong and Liu, Chang and Xu, Cheng and Li, Chushan and Gu, Yu and Watanabe, Kenji and Taniguchi, Takashi and Tong, Bingbing and Jia, Jinfeng and Shi, Zhiwen and Jiang, Shengwei and Zhang, Yang and Liu, Xiaoxue and Li, Tingxin},
  journal = {Phys. Rev. X},
  volume = {13},
  issue = {3},
  pages = {031037},
  numpages = {12},
  year = {2023},
  month = {Sep},
  publisher = {American Physical Society},
  doi = {10.1103/PhysRevX.13.031037},
  url = {https://link.aps.org/doi/10.1103/PhysRevX.13.031037}
}

@article{ji2024local,
  title={Local probe of bulk and edge states in a fractional {Chern} insulator},
  author={Ji, Zhurun and Park, Heonjoon and Barber, Mark E and Hu, Chaowei and Watanabe, Kenji and Taniguchi, Takashi and Chu, Jiun-Haw and Xu, Xiaodong and Shen, Zhi-Xun},
  journal={Nature},
  volume={635},
  number={8039},
  pages={578--583},
  year={2024},
  publisher={Nature Publishing Group UK London}
}

@article{redekop2024direct,
  title={Direct magnetic imaging of fractional Chern insulators in twisted MoTe $ \_2 $ with a superconducting sensor},
  author={Redekop, Evgeny and Zhang, Canxun and Park, Heonjoon and Cai, Jiaqi and Anderson, Eric and Sheekey, Owen and Arp, Trevor and Babikyan, Grigory and Salters, Samuel and Watanabe, Kenji and others},
  journal={arXiv preprint arXiv:2405.10269},
  year={2024}
}

@article{xu2025signatures,
  title={Signatures of unconventional superconductivity near reentrant and fractional quantum anomalous {Hall} insulators},
  author={Xu, Fan and Sun, Zheng and Li, Jiayi and Zheng, Ce and Xu, Cheng and Gao, Jingjing and Jia, Tongtong and Watanabe, Kenji and Taniguchi, Takashi and Tong, Bingbing and others},
  journal={arXiv preprint arXiv:2504.06972},
  year={2025}
}

@article{abouelkomsan2020,
  title = {Particle-Hole Duality, Emergent {Fermi} Liquids, and Fractional {Chern} Insulators in Moir\'e Flatbands},
  author = {Abouelkomsan, Ahmed and Liu, Zhao and Bergholtz, Emil J.},
  journal = {Phys. Rev. Lett.},
  volume = {124},
  issue = {10},
  pages = {106803},
  numpages = {6},
  year = {2020},
  month = {Mar},
  publisher = {American Physical Society},
  doi = {10.1103/PhysRevLett.124.106803},
  url = {https://link.aps.org/doi/10.1103/PhysRevLett.124.106803}
}

@article{repellin2020,
  title = {Chern bands of twisted bilayer graphene: Fractional {Chern} insulators and spin phase transition},
  author = {Repellin, C\'ecile and Senthil, T.},
  journal = {Phys. Rev. Res.},
  volume = {2},
  issue = {2},
  pages = {023238},
  numpages = {8},
  year = {2020},
  month = {May},
  publisher = {American Physical Society},
  doi = {10.1103/PhysRevResearch.2.023238},
  url = {https://link.aps.org/doi/10.1103/PhysRevResearch.2.023238}
}

@article{liu2021,
  title = {Gate-Tunable Fractional {Chern} Insulators in Twisted Double Bilayer Graphene},
  author = {Liu, Zhao and Abouelkomsan, Ahmed and Bergholtz, Emil J.},
  journal = {Phys. Rev. Lett.},
  volume = {126},
  issue = {2},
  pages = {026801},
  numpages = {6},
  year = {2021},
  month = {Jan},
  publisher = {American Physical Society},
  doi = {10.1103/PhysRevLett.126.026801},
  url = {https://link.aps.org/doi/10.1103/PhysRevLett.126.026801}
}

@article{li2021,
  title = {Spontaneous fractional {Chern} insulators in transition metal dichalcogenide moir\'e superlattices},
  author = {Li, Heqiu and Kumar, Umesh and Sun, Kai and Lin, Shi-Zeng},
  journal = {Phys. Rev. Res.},
  volume = {3},
  issue = {3},
  pages = {L032070},
  numpages = {6},
  year = {2021},
  month = {Sep},
  publisher = {American Physical Society},
  doi = {10.1103/PhysRevResearch.3.L032070},
  url = {https://link.aps.org/doi/10.1103/PhysRevResearch.3.L032070}
}

@article{crepel2023,
  title = {Anomalous {Hall} metal and fractional {Chern} insulator in twisted transition metal dichalcogenides},
  author = {Cr\'epel, Valentin and Fu, Liang},
  journal = {Phys. Rev. B},
  volume = {107},
  issue = {20},
  pages = {L201109},
  numpages = {5},
  year = {2023},
  month = {May},
  publisher = {American Physical Society},
  doi = {10.1103/PhysRevB.107.L201109},
  url = {https://link.aps.org/doi/10.1103/PhysRevB.107.L201109}
}

@article{reddy2023fci,
  title = {Fractional quantum anomalous Hall states in twisted bilayer ${\mathrm{MoTe}}_{2}$ and ${\mathrm{WSe}}_{2}$},
  author = {Reddy, Aidan P. and Alsallom, Faisal and Zhang, Yang and Devakul, Trithep and Fu, Liang},
  journal = {Phys. Rev. B},
  volume = {108},
  issue = {8},
  pages = {085117},
  numpages = {10},
  year = {2023},
  month = {Aug},
  publisher = {American Physical Society},
  doi = {10.1103/PhysRevB.108.085117},
  url = {https://link.aps.org/doi/10.1103/PhysRevB.108.085117}
}

@article{fonseca2025gradient,
  title={Gradient-based search of quantum phases: discovering unconventional fractional {Chern} insulators},
  author={Fonseca, Andr{\'e} Grossi and Wang, Eric and Vaidya, Sachin and Ledwith, Patrick J and Vishwanath, Ashvin and Solja{\v{c}}i{\'c}, Marin},
  journal={arXiv preprint arXiv:2509.10438},
  year={2025},
  url={https://arxiv.org/abs/2509.10438}
}

@article{shi2025effects,
  title={Effects of {Berry} curvature on ideal band magnetorotons},
  author={Shi, Jingtian and Cano, Jennifer and Morales-Dur{\'a}n, Nicol{\'a}s},
  journal={arXiv preprint arXiv:2503.15900},
  year={2025},
  url={https://arxiv.org/abs/2503.15900}
}

@article{liu2024kan,
  title={{KAN: Kolmogorov-Arnold} networks},
  author={Liu, Ziming and Wang, Yixuan and Vaidya, Sachin and Ruehle, Fabian and Halverson, James and Solja{\v{c}}i{\'c}, Marin and Hou, Thomas Y and Tegmark, Max},
  journal={arXiv preprint arXiv:2404.19756},
  year={2024},
  url={https://arxiv.org/abs/2404.19756}
}

@Article{varjas2022,
	title={{Topological lattice models with constant Berry curvature}},
	author={Daniel Varjas and Ahmed Abouelkomsan and Kang Yang and Emil J. Bergholtz},
	journal={SciPost Phys.},
	volume={12},
	pages={118},
	year={2022},
	publisher={SciPost},
	doi={10.21468/SciPostPhys.12.4.118},
	url={https://scipost.org/10.21468/SciPostPhys.12.4.118},
}

@article{Mera2021b,
  title = {K\"ahler geometry and Chern insulators: Relations between topology and the quantum metric},
  author = {Mera, Bruno and Ozawa, Tomoki},
  journal = {Phys. Rev. B},
  volume = {104},
  issue = {4},
  pages = {045104},
  numpages = {13},
  year = {2021},
  month = {Jul},
  publisher = {American Physical Society},
  doi = {10.1103/PhysRevB.104.045104},
  url = {https://link.aps.org/doi/10.1103/PhysRevB.104.045104}
}

@article{Lee2017,
  title = {Band structure engineering of ideal fractional Chern insulators},
  author = {Lee, Ching Hua and Claassen, Martin and Thomale, Ronny},
  journal = {Phys. Rev. B},
  volume = {96},
  issue = {16},
  pages = {165150},
  numpages = {16},
  year = {2017},
  month = {Oct},
  publisher = {American Physical Society},
  doi = {10.1103/PhysRevB.96.165150},
  url = {https://link.aps.org/doi/10.1103/PhysRevB.96.165150}
}

@article{Wang2021,
  title = {Exact Landau Level Description of Geometry and Interaction in a Flatband},
  author = {Wang, Jie and Cano, Jennifer and Millis, Andrew J. and Liu, Zhao and Yang, Bo},
  journal = {Phys. Rev. Lett.},
  volume = {127},
  issue = {24},
  pages = {246403},
  numpages = {6},
  year = {2021},
  month = {Dec},
  publisher = {American Physical Society},
  doi = {10.1103/PhysRevLett.127.246403},
  url = {https://link.aps.org/doi/10.1103/PhysRevLett.127.246403}
}

@article{Ledwith2020,
  title = {Fractional Chern insulator states in twisted bilayer graphene: An analytical approach},
  author = {Ledwith, Patrick J. and Tarnopolsky, Grigory and Khalaf, Eslam and Vishwanath, Ashvin},
  journal = {Phys. Rev. Res.},
  volume = {2},
  issue = {2},
  pages = {023237},
  numpages = {12},
  year = {2020},
  month = {May},
  publisher = {American Physical Society},
  doi = {10.1103/PhysRevResearch.2.023237},
  url = {https://link.aps.org/doi/10.1103/PhysRevResearch.2.023237}
}

@article{bergholtz_topological_2013,
	title = {Topological {Flat} {Band} {Models} and {Fractional} {Chern} {Insulators}},
	volume = {27},
	issn = {0217-9792, 1793-6578},
	url = {http://arxiv.org/abs/1308.0343},
	doi = {10.1142/S021797921330017X},
	number = {24},
	urldate = {2024-12-09},
	journal = {Int. J. Mod. Phys. B},
	author = {Bergholtz, Emil J. and Liu, Zhao},
	month = sep,
	year = {2013},
	pages = {1330017},
}

@article{jackson_geometric_2015,
	title = {Geometric stability of topological lattice phases},
	volume = {6},
	url = {https://www.nature.com/articles/ncomms9629},
	number = {1},
	journal = {Nat Commun},
	author = {Jackson, T. S. and Möller, Gunnar and Roy, Rahul},
	year = {2015},
	pages = {8629},
}

@article{ledwith_vortexability_2023,
  title = {Vortexability: A unifying criterion for ideal fractional {Chern} insulators},
  author = {Ledwith, Patrick J. and Vishwanath, Ashvin and Parker, Daniel E.},
  journal = {Phys. Rev. B},
  volume = {108},
  issue = {20},
  pages = {205144},
  numpages = {20},
  year = {2023},
  month = {Nov},
  publisher = {American Physical Society},
  doi = {10.1103/PhysRevB.108.205144},
  url = {https://link.aps.org/doi/10.1103/PhysRevB.108.205144}
}

@article{tang_high-temperature_2011,
	title = {High-{Temperature} {Fractional} {Quantum} {Hall} {States}},
	volume = {106},
	copyright = {http://link.aps.org/licenses/aps-default-license},
	issn = {0031-9007, 1079-7114},
	url = {https://link.aps.org/doi/10.1103/PhysRevLett.106.236802},
	doi = {10.1103/PhysRevLett.106.236802},
	number = {23},
	urldate = {2024-12-11},
	journal = {Phys. Rev. Lett.},
	author = {Tang, Evelyn and Mei, Jia-Wei and Wen, Xiao-Gang},
	month = jun,
	year = {2011},
	pages = {236802},
}

@article{wu_zoology_2012,
	title = {Zoology of fractional {Chern} insulators},
	volume = {85},
	copyright = {http://link.aps.org/licenses/aps-default-license},
	issn = {1098-0121, 1550-235X},
	url = {https://link.aps.org/doi/10.1103/PhysRevB.85.075116},
	doi = {10.1103/PhysRevB.85.075116},
	number = {7},
	urldate = {2024-12-11},
	journal = {Phys. Rev. B},
	author = {Wu, Yang-Le and Bernevig, B. Andrei and Regnault, N.},
	month = feb,
	year = {2012},
	pages = {075116},
}

@article{simon_fractional_2015,
	title = {Fractional {Chern} insulators in bands with zero {Berry} curvature},
	volume = {92},
	copyright = {http://link.aps.org/licenses/aps-default-license},
	issn = {1098-0121, 1550-235X},
	url = {https://link.aps.org/doi/10.1103/PhysRevB.92.195104},
	doi = {10.1103/PhysRevB.92.195104},
	number = {19},
	urldate = {2025-03-27},
	journal = {Phys. Rev. B},
	author = {Simon, Steven H. and Harper, Fenner and Read, N.},
	month = nov,
	year = {2015},
	pages = {195104},
}

@article{wang_fractional_2011,
	title = {Fractional {Quantum} {Hall} {Effect} of {Hard}-{Core} {Bosons} in {Topological} {Flat} {Bands}},
	volume = {107},
	copyright = {http://link.aps.org/licenses/aps-default-license},
	issn = {0031-9007, 1079-7114},
	url = {https://link.aps.org/doi/10.1103/PhysRevLett.107.146803},
	doi = {10.1103/PhysRevLett.107.146803},
	number = {14},
	urldate = {2025-03-31},
	journal = {Phys. Rev. Lett.},
	author = {Wang, Yi-Fei and Gu, Zheng-Cheng and Gong, Chang-De and Sheng, D. N.},
	month = sep,
	year = {2011},
	pages = {146803},
}

@article{pokrovskySimpleModelFractional1985,
  title = {A Simple Model for Fractional {{Hall}} Effect},
  author = {Pokrovsky, V. L. and Talapov, A. L.},
  year = {1985},
  month = {aug},
  journal = {Journal of Physics C: Solid State Physics},
  volume = {18},
  number = {23},
  pages = {L691},
  issn = {0022-3719},
  doi = {10.1088/0022-3719/18/23/002},
  urldate = {2024-08-16},
  langid = {english},
}

@article{hafezi_fractional_2007,
	title = {Fractional quantum {Hall} effect in optical lattices},
	volume = {76},
	url = {https://link.aps.org/doi/10.1103/PhysRevA.76.023613},
	doi = {10.1103/PhysRevA.76.023613},
	number = {2},
	urldate = {2025-05-14},
	journal = {Phys. Rev. A},
	author = {Hafezi, M. and Sørensen, A. S. and Demler, E. and Lukin, M. D.},
	month = aug,
	year = {2007},
	pages = {023613},
}

@incollection{liu_recent_2024,
	title = {Recent {Developments} in {Fractional} {Chern} {Insulators}},
	url = {http://arxiv.org/abs/2208.08449},
	urldate = {2025-05-15},
	author = {Liu, Zhao and Bergholtz, Emil J.},
	year = {2024},
	doi = {10.1016/B978-0-323-90800-9.00136-0},
	note = {arXiv:2208.08449},
	pages = {515--538}
}

@article{yang_singular_2025,
  title = {Fractional Quantum Anomalous {Hall} Effect in a Singular Flat Band},
  author = {Yang, Wenqi and Zhai, Dawei and Tan, Tixuan and Fan, Feng-Ren and Lin, Zuzhang and Yao, Wang},
  journal = {Phys. Rev. Lett.},
  volume = {134},
  issue = {19},
  pages = {196501},
  numpages = {7},
  year = {2025},
  month = {May},
  publisher = {American Physical Society},
  doi = {10.1103/PhysRevLett.134.196501},
  url = {https://link.aps.org/doi/10.1103/PhysRevLett.134.196501}
}

@article{lin_fractional_2025,
	title = {Fractional {Chern} insulator states in an isolated flat band of zero {Chern} number},
	url = {http://arxiv.org/abs/2505.09009},
	publisher = {arXiv},
	author = {Lin, Zuzhang and Yang, Wenqi and Lu, Hongyu and Zhai, Dawei and Yao, Wang},
	month = may,
	year = {2025},
	journal = {arXiv:2505.09009},
}

@article{neupert_fci_2011,
  title = {Fractional Quantum {Hall} States at Zero Magnetic Field},
  author = {Neupert, Titus and Santos, Luiz and Chamon, Claudio and Mudry, Christopher},
  journal = {Phys. Rev. Lett.},
  volume = {106},
  issue = {23},
  pages = {236804},
  numpages = {4},
  year = {2011},
  month = {Jun},
  publisher = {American Physical Society},
  doi = {10.1103/PhysRevLett.106.236804},
  url = {https://link.aps.org/doi/10.1103/PhysRevLett.106.236804}
}

@article{laughlin_anomalous_1983,
	title = {Anomalous {Quantum} {Hall} {Effect}: {An} {Incompressible} {Quantum} {Fluid} with {Fractionally} {Charged} {Excitations}},
	volume = {50},
	url = {https://link.aps.org/doi/10.1103/PhysRevLett.50.1395},
	doi = {10.1103/PhysRevLett.50.1395},
	number = {18},
	journal = {Phys. Rev. Lett.},
	author = {Laughlin, R. B.},
	year = {1983},
	pages = {1395--1398}
}

@article{oshikawa_commens_2000,
  title = {Commensurability, Excitation Gap, and Topology in Quantum Many-Particle Systems on a Periodic Lattice},
  author = {Oshikawa, Masaki},
  journal = {Phys. Rev. Lett.},
  volume = {84},
  issue = {7},
  pages = {1535--1538},
  numpages = {0},
  year = {2000},
  month = {Feb},
  publisher = {American Physical Society},
  doi = {10.1103/PhysRevLett.84.1535},
  url = {https://link.aps.org/doi/10.1103/PhysRevLett.84.1535}
}

@article{bernevig_emergent_2012,
	title = {Emergent many-body translational symmetries of {Abelian} and non-{Abelian} fractionally filled topological insulators},
	volume = {85},
	url = {https://link.aps.org/doi/10.1103/PhysRevB.85.075128},
	number = {7},
	journal = {Phys. Rev. B},
	author = {Bernevig, B. Andrei and Regnault, N.},
	month = feb,
	year = {2012},
	pages = {075128},
}

@article{park_observation_2023,
	title = {Observation of fractionally quantized anomalous {Hall} effect},
	volume = {622},
	copyright = {2023 The Author(s), under exclusive licence to Springer Nature Limited},
	issn = {1476-4687},
	url = {https://www.nature.com/articles/s41586-023-06536-0},
	doi = {10.1038/s41586-023-06536-0},
	number = {7981},
	urldate = {2025-06-07},
	journal = {Nature},
	author = {Park, Heonjoon and Cai, Jiaqi and Anderson, Eric and Zhang, Yinong and Zhu, Jiayi and Liu, Xiaoyu and Wang, Chong and Holtzmann, William and Hu, Chaowei and Liu, Zhaoyu and Taniguchi, Takashi and Watanabe, Kenji and Chu, Jiun-Haw and Cao, Ting and Fu, Liang and Yao, Wang and Chang, Cui-Zu and Cobden, David and Xiao, Di and Xu, Xiaodong},
	month = oct,
	year = {2023},
	pages = {74--79},
}

@article{lu_fractional_2024,
	title = {Fractional quantum anomalous {Hall} effect in multilayer graphene},
	volume = {626},
	copyright = {2024 The Author(s), under exclusive licence to Springer Nature Limited},
	issn = {1476-4687},
	url = {https://www.nature.com/articles/s41586-023-07010-7},
	doi = {10.1038/s41586-023-07010-7},
	number = {8000},
	urldate = {2025-06-07},
	journal = {Nature},
	author = {Lu, Zhengguang and Han, Tonghang and Yao, Yuxuan and Reddy, Aidan P. and Yang, Jixiang and Seo, Junseok and Watanabe, Kenji and Taniguchi, Takashi and Fu, Liang and Ju, Long},
	month = feb,
	year = {2024},
	pages = {759--764},
}

@article{roy_band_2014,
	title = {Band geometry of fractional topological insulators},
	volume = {90},
	url = {https://link.aps.org/doi/10.1103/PhysRevB.90.165139},
	doi = {10.1103/PhysRevB.90.165139},
	number = {16},
	journal = {Phys. Rev. B},
	author = {Roy, Rahul},
	month = oct,
	year = {2014},
	pages = {165139},
}

@article{girvin_magneto-roton_1986,
	title = {Magneto-roton theory of collective excitations in the fractional quantum {Hall} effect},
	volume = {33},
	copyright = {http://link.aps.org/licenses/aps-default-license},
	issn = {0163-1829},
	url = {https://link.aps.org/doi/10.1103/PhysRevB.33.2481},
	doi = {10.1103/PhysRevB.33.2481},
	number = {4},
	urldate = {2025-06-07},
	journal = {Phys. Rev. B},
	author = {Girvin, S. M. and MacDonald, A. H. and Platzman, P. M.},
	month = feb,
	year = {1986},
	pages = {2481--2494},
}

@article{ozawa_relations_2021,
  title = {Relations between topology and the quantum metric for {Chern} insulators},
  author = {Ozawa, Tomoki and Mera, Bruno},
  journal = {Phys. Rev. B},
  volume = {104},
  issue = {4},
  pages = {045103},
  numpages = {13},
  year = {2021},
  month = {Jul},
  publisher = {American Physical Society},
  doi = {10.1103/PhysRevB.104.045103},
  url = {https://link.aps.org/doi/10.1103/PhysRevB.104.045103}
}

@article{parameswaran2013fractional,
  title={Fractional quantum Hall physics in topological flat bands},
  author={Parameswaran, Siddharth A and Roy, Rahul and Sondhi, Shivaji L},
  journal={C. R. Phys.},
  volume={14},
  number={9-10},
  pages={816--839},
  year={2013},
  url = {https://comptes-rendus.academie-sciences.fr/physique/articles/10.1016/j.crhy.2013.04.003/}
}

@article{ardonne2008degeneracy,
  title={Degeneracy of non-{Abelian} quantum {Hall} states on the torus: domain walls and conformal field theory},
  author={Ardonne, Eddy and Bergholtz, Emil J and Kailasvuori, Janik and Wikberg, Emma},
  journal={J. Stat. Mech.},
  volume={2008},
  number={04},
  pages={P04016},
  year={2008},
  url={https://iopscience.iop.org/article/10.1088/1742-5468/2008/04/P04016}
}

@article{sun_2011_flatband,
  title = {Nearly Flatbands with Nontrivial Topology},
  author = {Sun, Kai and Gu, Zhengcheng and Katsura, Hosho and Das Sarma, S.},
  journal = {Phys. Rev. Lett.},
  volume = {106},
  issue = {23},
  pages = {236803},
  numpages = {4},
  year = {2011},
  month = {Jun},
  publisher = {American Physical Society},
  doi = {10.1103/PhysRevLett.106.236803},
  url = {https://link.aps.org/doi/10.1103/PhysRevLett.106.236803}
}

@article{spanton2018observation,
  title={Observation of fractional {Chern} insulators in a van der {Waals} heterostructure},
  author={Spanton, Eric M and Zibrov, Alexander A and Zhou, Haoxin and Taniguchi, Takashi and Watanabe, Kenji and Zaletel, Michael P and Young, Andrea F},
  journal={Science},
  volume={360},
  number={6384},
  pages={62--66},
  year={2018},
  publisher={American Association for the Advancement of Science},
  url={https://www.science.org/doi/10.1126/science.aan8458}
}

@article{xie2021fractional,
  title={Fractional {Chern} insulators in magic-angle twisted bilayer graphene},
  author={Xie, Yonglong and Pierce, Andrew T and Park, Jeong Min and Parker, Daniel E and Khalaf, Eslam and Ledwith, Patrick and Cao, Yuan and Lee, Seung Hwan and Chen, Shaowen and Forrester, Patrick R and others},
  journal={Nature},
  volume={600},
  number={7889},
  pages={439--443},
  year={2021},
  publisher={Nature Publishing Group UK London},
  url = {https://www.nature.com/articles/s41586-021-04002-3}
}

@article{zeng2023thermodynamic,
  title={Thermodynamic evidence of fractional Chern insulator in moir{\'e} MoTe2},
  author={Zeng, Yihang and Xia, Zhengchao and Kang, Kaifei and Zhu, Jiacheng and Kn{\"u}ppel, Patrick and Vaswani, Chirag and Watanabe, Kenji and Taniguchi, Takashi and Mak, Kin Fai and Shan, Jie},
  journal={Nature},
  volume={622},
  number={7981},
  pages={69--73},
  year={2023},
  publisher={Nature Publishing Group UK London},
  url={https://www.nature.com/articles/s41586-023-06452-3}
}

@article{wu2025modeling,
  title={Modeling Quantum Geometry for Fractional {Chern} Insulators with unsupervised learning},
  author={Wu, Ang-Kun and Primeau, Louis and Zhang, Jingtao and Sun, Kai and Zhang, Yang and Lin, Shi-Zeng},
  journal={arXiv:2510.03018},
  year={2025},
  url={https://arxiv.org/abs/2510.03018}
}
\end{document}